\documentstyle[12pt,epsf]{article} 

\def\lsim{\mathrel{\lower2.5pt\vbox{\lineskip=0pt\baselineskip=0pt 
           \hbox{$<$}\hbox{$\sim$}}}} 
\def\gsim{\mathrel{\lower2.5pt\vbox{\lineskip=0pt\baselineskip=0pt 
           \hbox{$>$}\hbox{$\sim$}}}} 

\def\L{\Lambda}
\def\o{\Omega}
\def\vp{\varphi}
\def\bl{\bar{\Lambda}}

\def\p{\partial}
\def\l{\lambda}
\def\k{\kappa}
\def\a{\alpha}

\begin{document} 
\begin{flushright}
DPNU-03-21\\ hep-th/0311006
\end{flushright}

\vspace{10mm}

\begin{center}
{\Large \bf 
 Scalar potential from de Sitter brane in 5D\\ and effective
 cosmological constant}

\vspace{20mm}
 Masato Ito 
 \footnote{E-mail address: mito@eken.phys.nagoya-u.ac.jp}
\end{center}

\begin{center}
{
\it 
{}Department of Physics, Nagoya University, Nagoya, 
JAPAN 464-8602 
}
\end{center}

\vspace{25mm}

\begin{abstract}
 We derive the scalar potential in zero mode effective action 
 arising from a de Sitter brane embedded in
 five dimensions with bulk cosmological constant $\L$.
 The scalar potential for a scalar field canonically normalized is
 given by the sum of exponential potentials.
 In the case of $\L=0$ and $\L>0$, we point out that
 the scalar potential has an unstable maximum at the origin
 and exponentially vanishes for large positive scalar field. 
 In the case of $\L<0$, the scalar potential has an unstable
 maximum at the origin and a local minimum.
 It is shown that the positive cosmological constant in $dS$ brane is reduced 
 by negative potential energy of scalar at minimum and that effective
 cosmological constant depends on a dimensionless quantity.
 Furthermore, we discuss the fate of our universe including the potential
 energy of the scalar.
\end{abstract} 
\newpage 
%
 \section{Introduction}
 \baselineskip=5.5mm

 An idea of braneworld that our world may be embedded in higher
 dimensional world is presently developing in particle physics as well
 as cosmology.
 This is because the braneworlds have several possibilities for explaining
 large hierarchy between electroweak scale and Planck scale,
 cosmological constant problem, inflation at early
 universe and accelerating universe at present, {\it etc}.
 Probably, the widespread braneworld started from the Randall-Sundrum
 model.
 An model proposed by Randall and Sundrum
 \cite{Randall:1999vf} assumes that the spacetime is non-factorizable
 with exponential warp factor and the extra space is non compact.
 The remarkable feature is that a zero mode gravity can be
 localized on a flat $3$-brane in $AdS_{5}$ background and
 the correction terms to the Newton potential are generated by the
 massive gravity with continuous modes.
 Moreover, the localization of gravity in the RS model have been analyzed
 in detail \cite{Lykken:1999nb,Giddings:2000mu,Csaki:2000fc}
 and the extension to more higher dimensions have been
 performed \cite{Ito:2001nc,Ito:2002tx,Ito:2001fd}. 

 It is known that the scalar potential arising from the braneworld
 can play an important role in cosmology and phenomenology.
 For instance, on searching mechanism for driving acceleration of
 our universe, it is expected that quintessence comes from the
 braneworld scenario.
 The possibility of quintessence in five dimensional
 dilatonic domain wall theory including RS model was investigated
 \cite{Myung:2000hk,Myung:2001sp}.
 The zero mode effective action with Einstein frame was studied in
 detail, unfortunately, it was shown that a scalar field canonically
 normalized cannot play the role of quintessence because of negative
 scalar potential.
 Furthermore, in the framework of string/M theory
 \cite{Kallosh:2001gr,Kallosh:2002gf} it was argued that
 the fate of our universe is determined by the properties of the scalar
 potential.
 Thus, when considering the scalar field as inflaton or quintessence in the
 cosmological context, it is important to study the properties of the
 scalar potential resulted from several models.
 
 In present paper, we consider the model of a de Sitter brane embedded in
 five dimensions with bulk cosmological constant $\L$
 \cite{Karch:2000ct,Ito:2002qp}.
 We extract the scalar potential from the four dimensional zero mode
 effective action depending on the sign of $\L$
 and the significant feature of the potential is studied in detail.
 In particular we are interested in the stability of the scalar
 potential and the effective cosmological constant in the brane. 
 This is because the fate of our universe is determined by the scalar potential.

 The paper is organized as follows.
 In section 2 we explain the action of the model, moreover,
 the metrics depending on sign of the bulk cosmological
 constant are presented. 
 In section 3 the four dimensional zero mode effective action is derived
 by performing integral of the extra dimension.
 We evaluate the scalar potential in the effective action and study the
 significant features of the potential depending on the sign of bulk
 cosmological constant.
 In section 4, the fate of our universe is discussed from the
 viewpoint of scalar cosmology.
 The conclusion is given in section 5.
 In appendix, we provide formulas of the integrals used in
 calculations of the effective action.
%
 \section{The model}

 We consider the model of a $dS$ brane embedded in five dimensions.
 We start by introducing the following action
 \cite{Karch:2000ct,Ito:2002qp} 
 \begin{eqnarray}
 S=\int d^{5}x\sqrt{-G}
    \left(\;\frac{\k^{-2}_{5}}{2}{\cal R}
    -\L\;\right)-\int d^{4}x\;\sqrt{-g}\;V\,,\label{eqn1}
 \end{eqnarray}
 where $\k^{2}_{5}$ is the five dimensional gravitational constant and
 $\L$ is the bulk cosmological constant, and $V$ is positive brane tension.
 We adopt the metric as follows
 \begin{eqnarray}
 ds^{2}=G_{MN}dx^{M}dx^{N}=\o^{-2}(z)
 \left(g_{\mu\nu}(x)dx^{\mu}dx^{\nu}+\vp^{2}(x)dz^{2}\right)\,,
 \label{eqn2}
 \end{eqnarray}
 where $g_{\mu\nu}$ is the four dimensional de Sitter metric with
 positive cosmological constant $\L_{4}$ and
 $z$-coordinate is the fifth direction with ${\bf Z}_{2}$ symmetry.
 Note that $\vp(x)$, diagonal component field for $z$ in the metric,
 corresponds to the graviscalar field from the four dimensional
 viewpoint \cite{Myung:2001sp}.
 Depending on the sign of $\L$, $\o(z)$ and $V$ can be expressed as
 \cite{Karch:2000ct,Ito:2002qp,Kaloper:1999sm}
 \begin{eqnarray}
 \L=0&:&\o(z)=e^{\sqrt{\bl}|z|}\,,\,
 V=6\k^{-2}_{5}\sqrt{\bl}\label{eqn3}\\
 && \nonumber\\
 \L>0&:&\o(z)=\frac{1}{L\sqrt{\bl}}
              \cosh\sqrt{\bl}(z_{0}+|z|)\,,\,
 V=\frac{6\k^{-2}_{5}}{L}\sqrt{L^{2}\bl-1}\label{eqn4}\\
 && \nonumber\\
 \L<0&:&\o(z)=\frac{1}{L\sqrt{\bl}}
              \sinh\sqrt{\bl}(z_{0}+|z|)\,,\,
 V=\frac{6\k^{-2}_{5}}{L}\sqrt{L^{2}\bl+1}\,,\label{eqn5}
 \end{eqnarray}
 where a $dS$ brane is located at $z=0$.
 Here we defined $3\bl\equiv \k^{2}_{4}\L_{4}$, where $\k^{2}_{4}$ is
 the four dimensional gravitational constant.
 Furthermore $L$ corresponding to the bulk curvature is given by
 \begin{eqnarray}
 L=\sqrt{\frac{6\k^{-2}_{5}}{|\L|}}\,.\label{eqn6}
 \end{eqnarray}
 In addition, $z_{0}$ is given by \cite{Karch:2000ct,Ito:2002qp}
 \begin{eqnarray}
 \L>0&:& z_{0}=\frac{1}{\bl}{\rm arc}\cosh
 \left(L\sqrt{\bl}\right)\label{eqn7}\\
 \L<0&:& z_{0}=\frac{1}{\bl}{\rm arc}\sinh
 \left(L\sqrt{\bl}\right)\label{eqn8}\,.
 \end{eqnarray}
 In the model presented here, it was shown that the four
 dimensional gravity can be produced at large distance due to
 localization of a normalizable zero mode in gravitational fluctuation
 on brane \cite{Karch:2000ct,Ito:2002qp,Kehagias:2002qk}.
%
 \section{Scalar potential}

 In this section, we calculate the four dimensional zero mode effective
 action for $\L=0$, $\L>0$ and $\L<0$, separately.
 Below, by using the appropriate conformal transformation,
 the effective action with Einstein frame and with
 a scalar field canonically normalized can be derived.
 Then the scalar potential is extracted from the action.

 Combining (\ref{eqn1}) with (\ref{eqn2}), the action is given by
 \begin{eqnarray}
 S =\int d^{4}x\sqrt{-g}L\,,\label{eqn9}
 \end{eqnarray}
 where
 \begin{eqnarray}
 L=\int^{\infty}_{-\infty}dz
 \left(\frac{\k^{-2}_{5}}{2}\o^{-3}\vp R
   +\k^{-2}_{5}\o^{-3}\left(4\frac{\o^{\prime\prime}}{\o}
                -10\frac{\o^{\prime 2}}{\o^{2}}\right)\vp^{-1}
   -\o^{-5}\L\vp\right)-V\,.\label{eqn10}
 \end{eqnarray}
 Here the prime denotes the derivative with respect to $z$.
 By performing the integral of $z$ in (\ref{eqn10}), we can obtain the
 four dimensional zero mode effective action.
 As indicated in \cite{Myung:2000hk,Myung:2001sp,Youm:2000vh},
 it turns out that the action corresponds to the
 Brans-Dicke model with zero Brans-Dicke parameter
 \cite{Dicke:1961gz,Brans:sx}.
 The significant feature of the scalar potential in the action
 will be discussed below.
 
\subsection{The case of $\L=0$}

 For $\L=0$, from (\ref{eqn3}), (\ref{eqn9}) and (\ref{eqn10}),
 the effective action is given by
 \begin{eqnarray}
 S=\int d^{4}x\sqrt{-g}
   \left(\frac{\k^{-2}_{5}}{3\sqrt{\bl}}\vp R
   +2\k^{-2}_{5}\sqrt{\bl}\left(2\vp^{-1}-3\right)\right)
 \label{eqn11}\,.
 \end{eqnarray}
 From the above action, we define the four dimensional gravitational
 constant as
 \begin{eqnarray}
 \k^{-2}_{4}=\frac{2\k^{-2}_{5}}{3\sqrt{\bl}}\,.\label{eqn12}
 \end{eqnarray}
 Thus the four dimensional gravitational constant is controlled by
 five dimensional gravitational constant and the cosmological constant
 in $dS$ brane.
 Consequently, we have
 \begin{eqnarray}
 S=\int d^{4}x\sqrt{-g}\left(
 \frac{\k^{-2}_{4}}{2}\vp R+3\k^{-2}_{4}\bl
 \left(2\vp^{-1}-3\right)
 \right)\,.\label{eqn13}
 \end{eqnarray}

 In order to obtain the action with Einstein frame and with a canonical
 kinetic term for scalar field, we need to perform the following conformal
 transformation
 \begin{eqnarray}
 g_{\mu\nu}=\vp^{-1}\bar{g}_{\mu\nu}\,,\;
 \vp=e^{\sqrt{2/3}\;\k_{4}\Phi}\,,\label{eqn14}
 \end{eqnarray}
 where a scalar field $\Phi$ is introduced.
 According to (\ref{eqn14}), the four dimensional zero mode effective
 action for a scalar field canonically normalized is given by
 \begin{eqnarray}
 S=\int d^{4}x\sqrt{-\bar{g}}
   \left(\frac{\k^{-2}_{4}}{2}\bar{\cal R}
   -\frac{1}{2}\left(\p\Phi\right)^{2}-V(\Phi)\right)\,,
 \label{eqn15}
 \end{eqnarray} 
 where the scalar potential $V$ with $\k_{4}=1$ can be expressed as
 \begin{eqnarray}
 V(\Phi)=3\bl\; e^{-2\sqrt{2/3}\;\Phi}
         \left(3-2 e^{-\sqrt{2/3}\;\Phi}\right)\,.
 \label{eqn16}
 \end{eqnarray}
%
\begin{figure}
      \epsfxsize=7cm
\center{\hspace{-5cm}$V/(3\bl)$\\
\vspace{0.3cm}
\begin{minipage}{7cm}\epsfbox{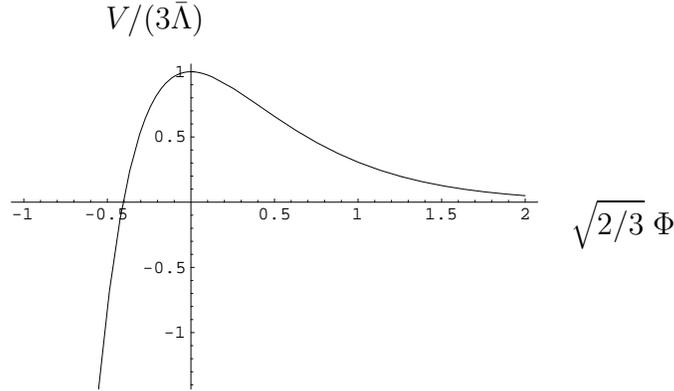}\end{minipage}
\hspace{0.3cm}$\sqrt{2/3}\;\Phi$}
\caption{The scalar potential $V(\Phi)$ in the case of $\L=0$.}
\label{fig1} 
\end{figure}
%
 Thus the scalar potential can be written by the sum of exponential
 potentials.
 The graph of $V(\Phi)$ is shown in Fig.\ref{fig1}.
 The potential has a maximum value $V_{\rm max}=3\bl$ at
 $\Phi=0$.
 It is obvious that we have positive potential for
 $e^{\sqrt{2/3}\;\Phi}>2/3$ and negative potential for
 $e^{\sqrt{2/3}\;\Phi}<2/3$.
 
 In the limit of $\Phi\rightarrow +\infty$, the asymptotic behavior is
 given by
 \begin{eqnarray}
 V(\Phi)\approx 9\bl\; e^{-2\sqrt{2/3}\;\Phi}\rightarrow
 +0\,.\label{eqn17}
 \end{eqnarray}
 For large positive scalar field, the potential exponentially approaches
 to zero.
 In the limit of $\Phi\rightarrow -\infty$,
 \begin{eqnarray}
 V(\Phi)\approx -6\bl\; e^{-3\sqrt{2/3}\;\Phi}\rightarrow
 -\infty\,.\label{eqn18}
 \end{eqnarray}
 Thus the potential goes to negative infinity for large negative scalar
 field.

 Note that the scalar potential has a maximum at the
 origin, and the effective mass at the maximum is
 tachyonic: $m^{2}=V^{\prime\prime}(0)=-12\bl$.
 As mentioned above, there is a zero minimum far away from the origin.

\subsection{The case of $\L>0$}

\begin{figure}
\centerline{
\begin{minipage}[b]{8cm}
\epsfxsize=7cm\epsfbox{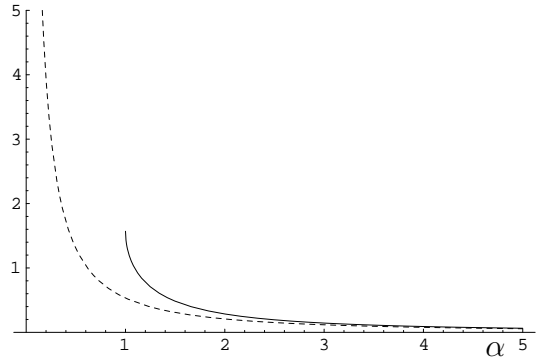}
\end{minipage}
\hspace{-0.8cm}$\a$
}
\caption{The graphs of $F(\a)$ (solid curve) and $G(\a)$ (broken curve).}
\label{fig2} 
\end{figure}

 For $\L>0$, from (\ref{eqn4}), (\ref{eqn7}) and (\ref{eqn10}), 
 the effective action can be evaluated by using
 (\ref{eqn37})$-$(\ref{eqn39}) in appendix.
 Therefore we have
 \begin{eqnarray}
 S&=&\int d^{4}x\sqrt{-g}
 \left[\;
 \frac{1}{2}\k^{-2}_{5}L\a
 \left(\frac{\pi}{2}-\frac{\sqrt{\a-1}}{\a}
       -{\rm arc}\tan\sqrt{\a-1}\right)\vp R\right.\nonumber\\
 &&\hspace{2cm}+\frac{\k^{-2}_{5}}{L}\a^{2}
   \left(\frac{3}{4}\pi+3\frac{\sqrt{\a-1}}{\a^{2}}
   -\frac{3\sqrt{\a-1}}{2\a}
       -\frac{3}{2}{\rm arc}\tan\sqrt{\a-1}\right)\vp^{-1}\nonumber\\
 &&\hspace{2cm}-\L L\a^{2}
   \left(\frac{3}{8}\pi-\frac{\sqrt{\a-1}}{2\a^{2}}
   -\frac{3\sqrt{\a-1}}{4\a}
       -\frac{3}{4}{\rm arc}\tan\sqrt{\a-1}\right)\vp\nonumber\\
 &&\left.\hspace{6cm}
 -\frac{6\k^{-2}_{5}}{L}\sqrt{\a-1}\;
 \right]\,,\label{eqn19}
 \end{eqnarray}
 where we introduce a dimensionless quantity
 \begin{eqnarray} 
 \a=L^{2}\bl \,.\label{eqn20}
 \end{eqnarray}
 As shown in the action, it is natural that the four dimensional
 gravitational constant is defined by
 \begin{eqnarray}
 \k^{-2}_{4}=\k^{-2}_{5}L\a F\left(\a\right)\,,\label{eqn21}
 \end{eqnarray}
 where 
 \begin{eqnarray}
 F\left(\a\right)=\frac{\pi}{2}-\frac{\sqrt{\a-1}}{\a}
       -{\rm arc}\tan\sqrt{\a-1}\label{eqn22}
 \end{eqnarray}
 for $\a\geq 1$. 
 Note that $F(\a)$ is monotonically decreasing function as shown in
 Fig.\ref{fig2}, for example,
 $F(1)=\pi/2$ and $F(\infty)=0$.
 Using the transformation of (\ref{eqn14}), in similar to the case of
 $\L=0$, after troublesome calculation the scalar potential can be 
 extracted from the effective action with Einstein frame.
 Therefore we obtain 
 \begin{eqnarray}
 V(\Phi)&=& \L Le^{-\sqrt{2/3}\;\Phi}
 \left[\;
 \frac{3}{4}\a^{2} F(\a)-\frac{\sqrt{\a-1}}{2}
 +\sqrt{\a-1}\;e^{-\sqrt{2/3}\;\Phi}
 \right.\nonumber\\
 &&\hspace{4cm}-\left.
 \left(\frac{1}{4}\a^2 F(\a)+\frac{\sqrt{\a-1}}{2}\right)
 e^{-2\sqrt{2/3}\;\Phi}
 \;\right]\,.\label{eqn23}
 \end{eqnarray}
%
%
 Here we set $\k_{4}=1$.
 It turns out that the scalar potential can be written by
 sum of exponential potentials.
 Next, we shall study the asymptotic behaviors of the potential for
 $|\Phi|\rightarrow\infty$.

 In the limit of $\Phi\rightarrow +\infty$, the potential is
 approximated as
 \begin{eqnarray}
 V(\Phi)\approx \frac{3}{4}\L L\left(\a^{2}F(\a)-\frac{2}{3}\sqrt{\a-1}
 \right)e^{-\sqrt{2/3}\;\Phi}\,.\label{eqn24}
 \end{eqnarray}
 Note that the asymptotic behavior for $\Phi\rightarrow +\infty$ is 
 determined by the sign of coefficient in front of exponential factor.
 According to the inequality $\a^{2}F(\a)>\frac{2}{3}\sqrt{\a-1}$ for
 $\a\geq 1$, 
 it is clear that $V(\Phi)\rightarrow +0$ for large positive $\Phi$.
%
\begin{figure}
      \epsfxsize=7cm
\center{\hspace{-5cm}$V/(\L L)$\\
\vspace{0.3cm}
\begin{minipage}[c]{7cm}\epsfbox{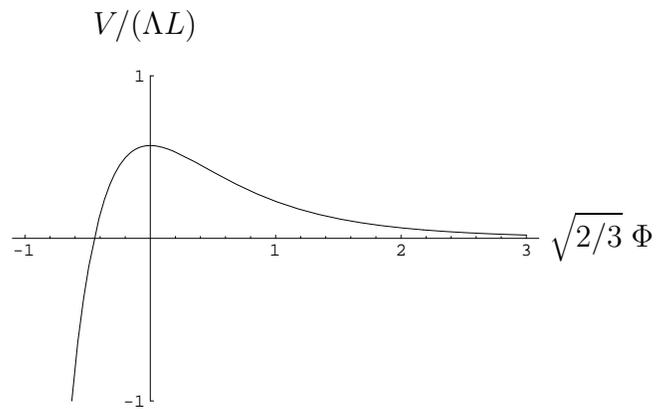}\end{minipage}
$\sqrt{2/3}\;\Phi$}
\caption{The scalar potential $V(\Phi)$ for $\L>0$ when $\a=2$.}
\label{fig3} 
\end{figure}
%
 In the limit of $\Phi\rightarrow -\infty$, the scalar potential is
 expressed as
 \begin{eqnarray}
 V(\Phi)\approx -\frac{\L L}{4}
 \left(\a^{2}F(\a)+2\sqrt{\a-1}\right)
 e^{-3\sqrt{2/3}\;\Phi}\rightarrow -\infty\,.\label{eqn25}
 \end{eqnarray}
 In order to clarify the behavior of scalar potential, 
 the graph of $V(\Phi)$ is shown in Fig.\ref{fig3} when $\a=2$.

 Examining the scalar potential of (\ref{eqn23}), the scalar potential
 has a maximum at $\Phi=0$,
 and we have $V_{\rm max}=\L L\a^{2} F(\a)/2$.
 Moreover the effective mass of scalar at the origin is tachyonic:
 $m^{2}=V^{\prime\prime}(0)=-\frac{2}{3}\k^{2}_{4}\L L\left(
 \sqrt{\a-1}+\frac{3}{2}\a^{2}F(\a)\right)$.  
 Note that the scalar potential exponentially vanishes for large
 positive scalar field.
 This implies the existence of a minimum away far from the origin.

\subsection{The case of $\L<0$}

 For $\L<0$, from (\ref{eqn5}), (\ref{eqn8}) and (\ref{eqn10}), 
 the effective action is evaluated by using
 (\ref{eqn40})$-$(\ref{eqn42}) in appendix.
 Consequently, we have
 \begin{eqnarray}
 S&=&\int d^{4}x\sqrt{-g}
 \left[\;
 \frac{1}{2}\k^{-2}_{5}L\a
 \left(\frac{\sqrt{1+\a}}{\a}+\log\frac{\sqrt{1+\a}-1}
 {\sqrt{\a}}\right)\vp R
 \right.\nonumber\\
 &&\hspace{2cm}+\frac{\k^{-2}_{5}}{L}\a^{2}\left(
    3\frac{\sqrt{1+\a}}{\a^{2}}+\frac{3\sqrt{1+\a}}{2\a}
    +\frac{3}{2}\log\frac{\sqrt{1+\a}-1}{\sqrt{\a}}
    \right)\vp^{-1}\nonumber\\
 &&\hspace{2cm}+|\L| L\a^{2}
   \left(
    \frac{\sqrt{1+\a}}{2\a^{2}}-\frac{3\sqrt{1+\a}}{4\a}
    -\frac{3}{4}\log\frac{\sqrt{1+\a}-1}{\sqrt{\a}}
    \right)\vp\nonumber\\
 &&\left.\hspace{6cm}
   -\frac{6\k^{-2}_{5}}{L}\sqrt{1+\a}\;\right]\,.
 \label{eqn26}
 \end{eqnarray}
 In this case we define the four dimensional gravitational constant as
 \begin{eqnarray}
 \k^{-2}_{4}=\k^{-2}_{5}L\a G\left(\a\right)\,,\label{eqn27}
 \end{eqnarray}
 where
 \begin{eqnarray}
 G\left(\a\right)=\frac{\sqrt{1+\a}}{\a}+\log\frac{\sqrt{1+\a}-1}
 {\sqrt{\a}}\,,\label{eqn28}
 \end{eqnarray}
 for $\a\geq 0$.
 As shown in Fig.\ref{fig2}, $G(\a)$ is monotonically decreasing function.
 It is obvious that the action of (\ref{eqn26}) with
 $\bl\rightarrow 0\;(\a\rightarrow 0)$ corresponding to a flat brane is
 equivalent to the effective action resulted from the RS model
 \cite{Myung:2000hk,Myung:2001sp,Youm:2000vh}.
 
 The transformation of (\ref{eqn14}) leads to the effective action
 with Einstein frame and with a scalar field canonically normalized, 
 consequently, we can obtain the scalar potential as follows
 \begin{eqnarray}
 V(\Phi)&=& |\L| Le^{-\sqrt{2/3}\;\Phi}
 \left[\;
 \frac{3}{4}\a^{2} G(\a)-\frac{\sqrt{\a+1}}{2}
 +\sqrt{\a+1}\;e^{-\sqrt{2/3}\;\Phi}
 \right.\nonumber\\
 &&\hspace{4cm}-\left.
 \left(\frac{1}{4}\a^{2}G(\a)+\frac{\sqrt{\a+1}}{2}\right)
 e^{-2\sqrt{2/3}\;\Phi}
 \;\right]\,.\label{eqn29}
 \end{eqnarray}
 Here we set $\k_{4}=1$.
 Comparing (\ref{eqn29}) with (\ref{eqn23}), 
 (\ref{eqn29}) is consistent with the equation
 with replacements of $F\rightarrow G$ and 
 $\sqrt{\a-1}\rightarrow\sqrt{\a+1}$ in (\ref{eqn23}). 
%
\begin{figure}
\center{\hspace{-12cm}$V/(|\L|L)$\\
\vspace{0.3cm}
\begin{minipage}{7cm}\epsfxsize=7cm\epsfbox{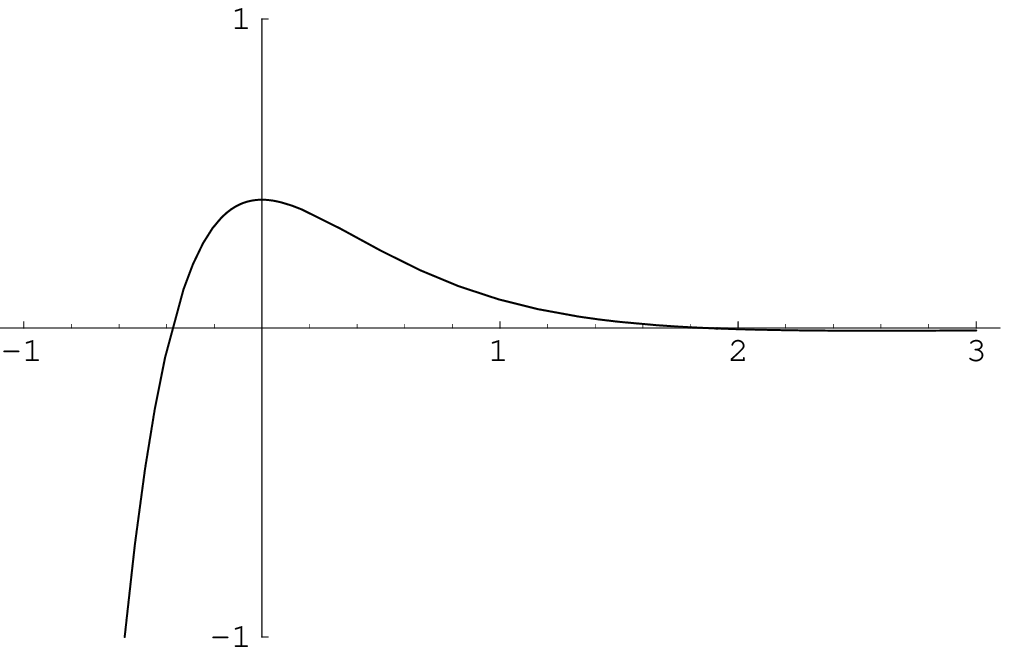}\end{minipage}
\hspace{0.5cm}
\begin{minipage}{7cm}\epsfxsize=6.7cm\epsfbox{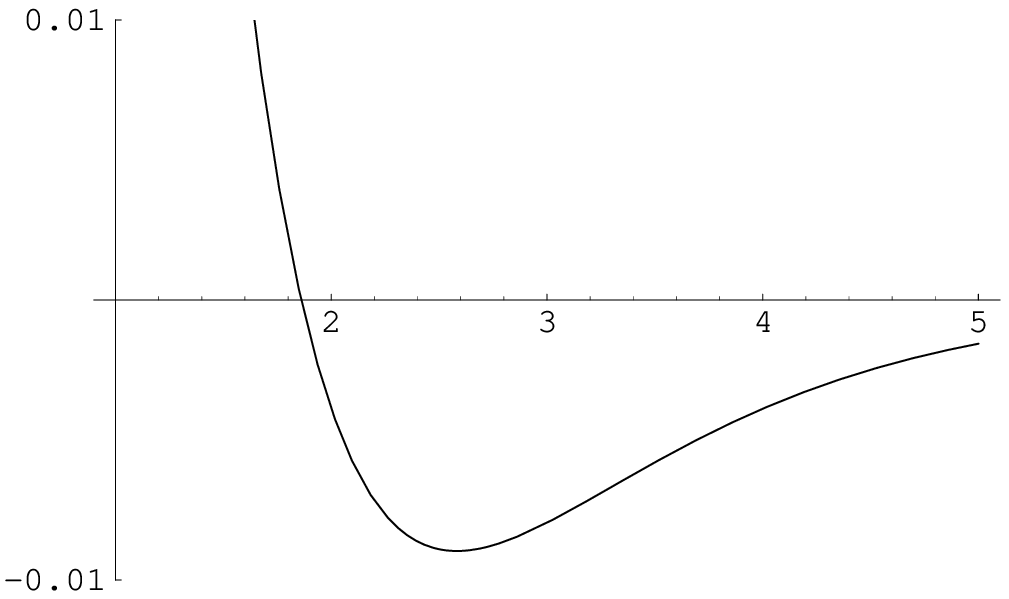}\end{minipage}
}
\caption{The scalar potential $V(\Phi)$ for $\L<0$ when $\a=2$.
The horizontal line is $x=\sqrt{2/3}\;\Phi$,
$-1<x<3$ (left) and  $1<x<5$ (right).}
\label{fig4} 
\end{figure}
%

 In the limit of $\Phi\rightarrow +\infty$, the potential is given by
 \begin{eqnarray}
 V(\Phi)\approx \frac{3}{4}|\L| L\left(\a^{2}G(\a)-\frac{2}{3}\sqrt{1+\a}
 \right)e^{-\sqrt{2/3}\;\Phi}\,.\label{eqn30}
 \end{eqnarray}
 From inequality $\a^{2}G(\a)<\frac{2}{3}\sqrt{1+\a}$ for $\a\geq 0$,
 it turns out that $V(\Phi)\rightarrow -0$ for $\Phi\rightarrow +\infty$.
 Taking the limit of $\Phi\rightarrow -\infty$, we have
 \begin{eqnarray}
 V(\Phi)\approx -\frac{|\L| L}{4}
 \left(\a^{2}G(\a)+2\sqrt{1+\a}\right)
 e^{-3\sqrt{2/3}\;\Phi}\rightarrow -\infty
 \,.\label{eqn31}
 \end{eqnarray}
 The graph of the scalar potential is shown in
 Fig.\ref{fig4} when $\a=2$.

 Examining (\ref{eqn29}), it turns out that the scalar potential has two
 extremum at $\Phi=0,\Phi_{0}$.
 At $\Phi=0$, there is a maximum with tachyonic mass:
 \begin{eqnarray}
 m^{2}=-\frac{2}{3}\k^{2}_{4}|\L|L\left(
 \sqrt{\a+1}+\frac{3}{2}\a^{2}G(\a)
 \right)\,.\label{32}
 \end{eqnarray}
 Furthermore, there exists a local minimum at $\Phi_{0}$:
 \begin{eqnarray}
 \Phi_{0}=\sqrt{\frac{3}{2}}\log\left(
 3\frac{2\sqrt{\a+1}+\a^{2}G(\a)}{2\sqrt{\a+1}-3\a^{2}G(\a)}
 \right)\,.\label{eqn33}
 \end{eqnarray}
 Note that the value of the minimum at $\Phi_{0}$ is always negative and
 the effective mass is given by
 \begin{eqnarray}
 V^{\prime\prime}(\Phi_{0})=\k^{2}_{4}|\L|Le^{-\sqrt{6}\;\Phi_{0}}\;
 \frac{2\sqrt{\a+1}+3\a^{2}G(\a)}{2\sqrt{\a+1}-3\a^{2}G(\a)}
 \left(2\sqrt{\a+1}+\a^{2}G(\a)\right)>0\,.\label{eqn34}
 \end{eqnarray}
 Accordingly this implies a local minimum at $\Phi_{0}$.
 The positive cosmological constant in $dS$ brane is influenced by
 contribution of potential energy of scalar field at minimum, namely, 
 the effective cosmological constant is given  by
 $\L_{\rm eff}=\L_{4}+V(\Phi_{0})$.
 Therefore we obtain
 \begin{eqnarray}
 \L_{\rm eff}=\frac{|\L|L}{2}
 \left(
 \a^{2}G(\a)-\frac{4\sqrt{\a+1}+3\a^{2}G(\a)}{27}
 \left(\frac{2\sqrt{\a+1}-3\a^{2}G(\a)}{4\sqrt{\a+1}+\a^{2}G(\a)}\right)^{2}
 \right)\,.\label{eqn35}
 \end{eqnarray}
 The second term corresponds to the potential energy of scalar field
 at minimum.
 Thus the positive cosmological constant in $dS$ brane is reduced by
 dynamics of scalar field.
 Estimating (\ref{eqn35}), $\L_{\rm eff}$ is monotonically
 increasing function for $\a$.
 Note that $\L_{\rm eff}<0$ for $0\leq \a\lsim 0.04$ and
 $\L_{\rm eff}>0$ for $0.04<\a$.
 Thus it is considered that the value of the effective cosmological
 constant in $dS$ brane can be tuned via scalar dynamics on the
 $dS$ brane.  
\section{Cosmological implications}

We would like to discuss the fate of the universe in the
framework of this model, assuming that our universe is dominated by
a scalar field $\Phi$.
It is natural to consider a scalar $\Phi$ as the component of dark
energy driving acceleration of the universe at present.

Here we review the important role of the scalar potential in
cosmological model \cite{Kallosh:2003mt}.
Particularly, two types are presented: I) The scalar potential $V$ slowly decreases
from positive to zero as a scalar rolls to infinity (slow-roll).
Assuming that dark energy corresponds to the energy of a slowly rolling
scalar with equation of state $(p_{D}\sim -\rho_{D})$, the universe
goes from dS regime (the present stage of acceleration) to Minkowski regime. 
In the case of exponential potential, Quintessence requires $V(\phi)\sim
e^{-\l\phi}$, where $\l < \sqrt{2}$.
II) The scalar potential $V$ has a negative minimum or it falls to the
negative infinity (free-fall).
In this case the fate of the universe reaches the stage of collapse
\cite{Felder:2002jk}. 
Below we apply the present model to the above two types.

The total potential energy in the $dS$ brane is expressed as
$V_{\rm tot}=\L_{4}+V(\Phi)$, where $\L_{4}$ is the cosmological constant in
$dS$ brane, assuming that $\L_{4}$ is a very small positive value.
In this paper the origin of fixing the magnitude of $\L_{4}$ isn't discussed.
Namely the plots of $V(\Phi)$ in Fig 1,3,4 are lifted by $\L_{4}$.  
Investigating total potential energy $V_{\rm tot}$ in the present model,
we obtain the results as follows.

i) For $\L=0$ the potential has both properties of slow-roll and free-fall as
shown in Fig.1.
The roll from the top to $\Phi=+\infty$ slowly decreases, however,
acceleration cannot arise because of 
$V_{\rm tot}\simeq e^{-\l\Phi}$ with $\l=2\sqrt{2/3}>\sqrt{2}$ from (\ref{eqn17}). 
The roll to $\Phi<0$ is free to fall to $V_{\rm tot}=-\infty$, namely, it implies
collapse of the universe.

ii) For $\L>0$, as shown in Fig.3, the potential has both properties of
slow-roll and free-fall.
The roll from the top to $\Phi=+\infty$ can generate acceleration because of 
$V_{\rm tot}\simeq e^{-\l\Phi}$ with $\l=\sqrt{2/3}<\sqrt{2}$ from (\ref{eqn24}). 
While the roll from the top to $V_{\rm tot}=-\infty$ leads to collapse of the universe.

iii) For $\L<0$, from Fig.4, the potential has the property of free-fall and a minimum
obtained in (\ref{eqn35}).
The roll of free-fall implies collapse of the universe.
The rolling scalar from the top to a minimum or from $\Phi=+\infty$ to
a minimum would reach a minimum $\L_{\rm eff}$ of (\ref{eqn35}), however, 
in this case the fate of the universe depends on the sign of $\L_{\rm eff}$.
The case of negative $\L_{\rm eff}$ ($0<\a<0.04$) implies a negative minimum, namely,
the universe collapses.
Although the case of positive $\L_{\rm eff}$ ($\a>0.04$) leads to $dS$ regime during
present stage of acceleration, the scalar is generically tunneling into
negative infinity and the universe eventually goes to collapse.

\section{Conclusion}

 In this paper we have studied the scalar potential by evaluating the
 four dimensional zero mode effective action resulting from the model of
 a $dS$ brane embedded in five dimensions with bulk cosmological
 constant $\L$.
 The scalar potential is explored in the case of $\L=0$, $\L>0$ and
 $\L<0$, separately.
 We pointed out that the potential for a scalar field canonically
 normalized is given by the sum of exponential potentials and
 discussed the fate of our universe from the viewpoint of scalar dominated
 cosmology.

 For $\L=0$ and $\L>0$, the scalar potential has an unstable maximum at
 the origin, and both properties of free-fall and slow-roll are shown.
 In the case of $\L>0$, the scalar potential can be peculiarly suitable for
 Quintessence potential.

 For $\L<0$, the potential has an unstable maximum
 at the origin and a local minimum at $\Phi_{0}$.
 If we are living in the minimum, the positive cosmological constant in $dS$ brane can be
 reduced by negative potential energy of scalar field at minimum.
 Consequently it turned out that the effective cosmological constant in the brane
 depending on a dimensionless quantity $\alpha$ in (\ref{eqn20}) is tuned via the
 dynamics of scalar field.
 In this case the fate of the universe eventually would collapse via
 tunneling effect even if we are living in the local minimum (false vacuum).

 Advanced astronomical observations \cite{Perlmutter:1998np} indicate
 that our universe is accelerating today and that the cosmological
 constant has sufficiently small positive value. 
 In toy model presented here, the value of the cosmological constant can
 be controlled by a dimensionless quantity $\a$ so as to be consistent with
 the observable value.
 However this is not to say that we solve the cosmological constant
 problem. This is because the mechanism for determining the value of
 $\a$ is unknown.
 Thus, in the development of cosmology \cite{Langlois:2002bb}, we think
 that it is important to investigate scalar potential resulted from braneworld. 

 Finally we give some comments with regard to cosmological context.
 We have to study not only the form of the scalar potential, but also
 the scalar evolution in the model.
 By analyzing the scalar evolution, we expect that the fate of our
 universe can be discussed in detail.
 We will describe it elsewhere.

%
 \section*{Appendix: Evaluation of integrals including hyperbolic functions}
 In Appendix, we provide formulas of integral calculations
 performed in section 3.
 Below all integral constants are omitted.

 For $m,n\in {\bf Z}$, we define $I[m,n]$ by
 \begin{eqnarray}
 I[m,n]=\int\sinh^{m}x\;\cosh^{n}x\;dx\,.\label{eqn36}
 \end{eqnarray}
 
 In section 3, the zero mode effective action can be obtained by using
 formulas as follows
 \begin{eqnarray}
 I[0,-3]&=&\frac{\sinh x}{2\cosh^{2} x}+
         \frac{1}{2}{\rm arc}\tan\left(\sinh x\right)\,,\label{eqn37}\\
 I[0,-5]&=&\frac{\sinh x}{4\cosh^{4}x}+\frac{3\sinh x}{8\cosh^{2}x}
          +\frac{3}{8}{\rm arc}\tan\left(\sinh x\right)\,,\label{eqn38}\\
 I[2,-5]&=&-\frac{\sinh x}{4\cosh^{4}x}+\frac{\sinh x}{8\cosh^{2}x}
          +\frac{1}{8}{\rm arc}\tan\left(\sinh x\right)\,,\label{eqn39}\\
 I[-3,0]&=&-\frac{\cosh x}{2\sinh^{2} x}
           -\frac{1}{2}\log\left|\tanh\frac{x}{2}\right|\,,\label{eqn40}\\ 
 I[-5,0]&=&-\frac{\cosh x}{4\sinh^{4}x}
          +\frac{3\cosh x}{8\sinh^{2}x}
          +\frac{3}{8}\log\left|\tanh\frac{x}{2}\right|\,,\label{eqn41}\\
 I[-5,2]&=&-\frac{\cosh x}{4\sinh^{4}x}
          -\frac{\cosh x}{8\sinh^{2}x}
          -\frac{1}{8}\log\left|\tanh\frac{x}{2}\right|\,.\label{eqn42}
 \end{eqnarray}
%
%
 
%

\begin{thebibliography}{99}
%
 \bibitem{Randall:1999vf}
 L.~Randall and R.~Sundrum,
 ``An alternative to compactification,''
 Phys.\ Rev.\ Lett.\  {\bf 83}, 4690 (1999) [hep-th/9906064].
%
 \bibitem{Lykken:1999nb}
 J.~Lykken and L.~Randall,
 ``The shape of gravity,'' JHEP {\bf 0006}, 014 (2000) [hep-th/9908076].
%
 \bibitem{Giddings:2000mu}
 S.~B.~Giddings, E.~Katz and L.~Randall,
 ``Linearized gravity in brane backgrounds,'' JHEP {\bf 0003}, 023 (2000)
 [hep-th/0002091].
%
 \bibitem{Csaki:2000fc}
 C.~Csaki, J.~Erlich, T.~J.~Hollowood and Y.~Shirman,
 ``Universal aspects of gravity localized on thick branes,''
 Nucl.\ Phys.\ B {\bf 581}, 309 (2000)
 [arXiv:hep-th/0001033].
%
 \bibitem{Ito:2001nc}
 M.~Ito,
 ``Newton's law in braneworlds with an infinite extra dimension,''
 Phys.\ Lett.\ B {\bf 528}, 269 (2002) [hep-th/0112224].
%
 \bibitem{Ito:2002tx}
 M.~Ito,
 ``Correction terms to Newton law due to induced gravity in AdS  background,''
 Phys.\ Lett.\ B {\bf 554}, 180 (2003)
 [arXiv:hep-th/0211268].
%
 \bibitem{Ito:2001fd}
 M.~Ito,
 ``Warped geometry in higher dimensions with an orbifold extra dimension,''
 Phys.\ Rev.\ D {\bf 64}, 124021 (2001) [hep-th/0105186].
%
 \bibitem{Myung:2000hk}
 Y.~S.~Myung and H.~W.~Lee,
 ``Schwarzschild black hole in the dilatonic domain wall,''
 Phys.\ Rev.\ D {\bf 63}, 064034 (2001)
 [arXiv:hep-th/0001211].
%
 \bibitem{Myung:2001sp}
 Y.~S.~Myung,
 ``Quintessence and brane world scenarios,''
 Mod.\ Phys.\ Lett.\ A {\bf 16}, 2187 (2001)
 [arXiv:hep-th/0107065].
%
 \bibitem{Kallosh:2001gr}
 R.~Kallosh, A.~D.~Linde, S.~Prokushkin and M.~Shmakova,
 ``Gauged supergravities, \cite{Felder:2002jk}de Sitter space and cosmology,''
 Phys.\ Rev.\ D {\bf 65}, 105016 (2002)
 [arXiv:hep-th/0110089].
%
 \bibitem{Kallosh:2002gf}
 R.~Kallosh, A.~Linde, S.~Prokushkin and M.~Shmakova,
 ``Supergravity, dark energy and the fate of the universe,''
 Phys.\ Rev.\ D {\bf 66}, 123503 (2002)
 [arXiv:hep-th/0208156].
%
 \bibitem{Karch:2000ct}
 A.~Karch and L.~Randall,
 ``Locally localized gravity,'' JHEP {\bf 0105}, 008 (2001)
 [hep-th/0011156].
%
 \bibitem{Ito:2002qp}
 M.~Ito,
 ``Localized gravity on de Sitter brane in five dimensions,''
 Europhys. Lett. {\bf 64}, 295 (2003)
 [arXiv:hep-th/0204113].
%
 \bibitem{Kaloper:1999sm}
 N.~Kaloper,
 ``Bent domain walls as braneworlds,''
 Phys.\ Rev.\ D {\bf 60}, 123506 (1999)
 [arXiv:hep-th/9905210].
%
 \bibitem{Kehagias:2002qk}
 A.~Kehagias and K.~Tamvakis,\cite{Felder:2002jk}
 ``Graviton localization and Newton law for a dS(4) brane in 5D bulk,''
 Class.\ Quant.\ Grav.\  {\bf 19}, L185 (2002)
 [arXiv:hep-th/0205009].
%
 \bibitem{Youm:2000vh}
 D.~Youm,
 ``A note on solitons in brane worlds,''
 Phys.\ Rev.\ D {\bf 63}, 047503 (2001)
 [arXiv:hep-th/0001166].
%
 \bibitem{Dicke:1961gz}
 R.~H.~Dicke,
 ``Mach's Principle And Invariance Under Transformation Of Units,''
 Phys.\ Rev.\  {\bf 125}, 2163 (1962).
%
 \bibitem{Brans:sx}
 C.~Brans and .~H.~Dicke,
 ``Mach's Principle And A Relativistic Theory Of Gravitation,''
 Phys.\ Rev.\  {\bf 124} (1961) 925.
%
 \bibitem{Perlmutter:1998np}
 S.~Perlmutter {\it et al.}  [Supernova Cosmology Project Collaboration],
 ``Measurements of Omega and Lambda from 42 High-Redshift Supernovae,''
 Astrophys.\ J.\  {\bf 517}, 565 (1999) [astro-ph/9812133].
%
 \bibitem{Kallosh:2003mt}
 R.~Kallosh and A.~Linde,
 ``Dark energy and the fate of the universe,''
 JCAP {\bf 0302}, 002 (2003)[arXiv:astro-ph/0301087].\cite{Felder:2002jk}
%
 \bibitem{Felder:2002jk}
 G.~N.~Felder, A.~V.~Frolov, L.~Kofman and A.~V.~Linde,
 ``Cosmology with negative potentials,''
 Phys.\ Rev.\ D {\bf 66}, 023507 (2002)[arXiv:hep-th/0202017].
%
 \bibitem{Langlois:2002bb}
 D.~Langlois,
 ``Brane cosmology: An introduction,''
 Prog.\ Theor.\ Phys.\ Suppl.\  {\bf 148}, 181 (2003)
 [arXiv:hep-th/0209261].
%

\end{thebibliography}
\end{document}